\documentclass[10pt, conference]{IEEEtran}
\usepackage{listings}
\usepackage{booktabs}
\usepackage{tcolorbox}
\IEEEoverridecommandlockouts
\usepackage{cite}
\usepackage{amsmath,amssymb,amsfonts}
\usepackage{algorithmic}
\usepackage{graphicx}
\usepackage{textcomp}
\usepackage{xcolor}

\usepackage[normalem]{ulem}

\usepackage[small,compact]{titlesec}
\titlespacing{\section}{2pt}{2pt}{2pt}
\titlespacing{\subsection}{2pt}{2pt}{2pt}
\titlespacing{\subsubsection}{2pt}{2pt}{2pt}

\newcommand{\figname}{{Fig.}}

\begin{document}

\title{Paradigm-Based Automatic HDL Code Generation Using LLMs\\
{}
}

\author{
\IEEEauthorblockN{Wenhao Sun$^1$,
Bing Li$^2$,
Grace Li Zhang$^3$,
Xunzhao Yin$^4$,
Cheng Zhuo$^4$,
Ulf Schlichtmann$^1$}
\IEEEauthorblockA{$^1$Technical University of Munich (TUM), 
$^2$University of Siegen, 
$^3$TU Darmstadt, $^4$Zhejiang University
}
\IEEEauthorblockA{Email: \{wenhao.sun, ulf.schlichtmann\}@tum.de, bing.li@uni-siegen.de,  \\ grace.zhang@tu-darmstadt.de, \{xzyin1,czhuo\}@zju.edu.cn}
}

\maketitle

\label{sec:abstract}

\begin{abstract}
While large language models (LLMs) have demonstrated the ability to generate hardware description language (HDL) code for digital circuits, they still face the hallucination problem, which can result in the generation of incorrect HDL code or misinterpretation of specifications. In this work, we introduce a human-expert-inspired method to mitigate the hallucination of LLMs and enhance their performance in HDL code generation. We begin by constructing specialized paradigm blocks that consist of several steps designed to divide and conquer generation tasks, mirroring the design methodology of human experts. These steps include information extraction, human-like design flows, and the integration of external tools. LLMs are then instructed to classify the type of circuit in order to match it with the appropriate paradigm block and execute the block to generate the HDL codes. Additionally, we propose a two-phase workflow for multi-round generation, aimed at effectively improving the testbench pass rate of the generated HDL codes within a limited number of generation and verification rounds. Experimental results demonstrate that our method significantly enhances the functional correctness of the generated Verilog code.
\end{abstract}
\section{Introduction}\label{sec:introduction}

As Moore's Law slows down, there is an increasing demand for customized very-large-scale integration (VLSI) design. A key step in the hardware design process is writing hardware description language (HDL) code. However, HDL programming is time-consuming and labor-intensive, leading to a growing interest in automatic HDL code generation from specifications in natural language.

Among the various solutions for automatic HDL code generation, large language models (LLMs) represent one of the most promising approaches \cite{xu2024llm,9927393}. LLMs have achieved remarkable success in diverse fields, such as machine translation \cite{translation} and robot trajectory planning \cite{robot}. In software design, LLMs have demonstrated the ability to generate code for various programming languages \cite{codegen,autobench,hlsrepair,qiu2024correctbench}. In hardware design, generative models have also been employed to create designs \cite{survey}, and researchers have recognized the potential of LLMs for generating HDL code \cite{benchmark}.

\begin{figure}[t!]
    \centering
    \includegraphics[width=0.48\textwidth]{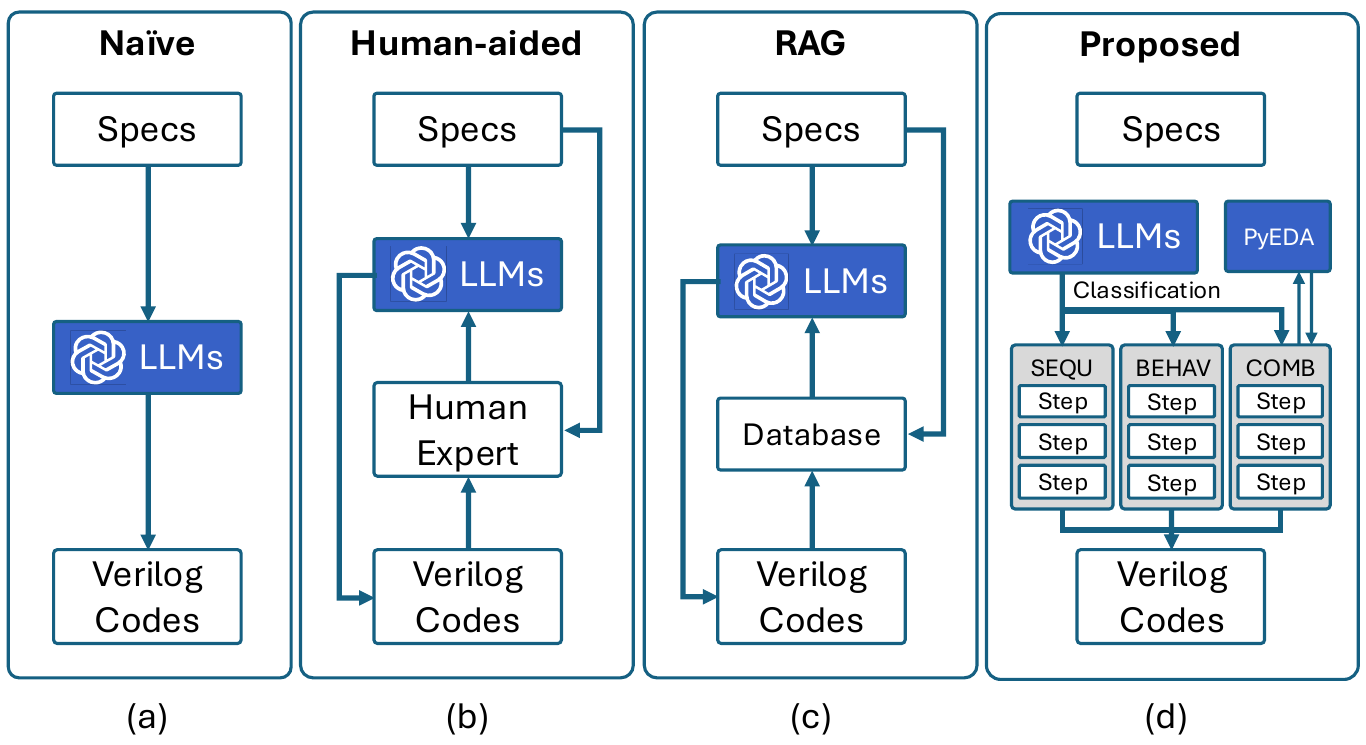}
    \caption{Comparison of Training-Free Methods: naïve generation, human-aided generation, and RAG generation are illustrated in (a), (b), and (c). By introducing the human-expert-inspired procedure, the proposed method illustrated in (d) enhances the performance of LLM code generation without the need for human labor or a database. 
    }
    \label{full}
\end{figure}

Despite the advantages of LLMs, several challenges hinder their application in HDL code generation. One significant challenge is the hallucination problem, which refers to the generation of plausible but incorrect code or a misunderstanding of specifications \cite{hallucination}. On one hand, this problem arises from the limited availability of training data in the HDL domain. On the other hand, the reasoning capabilities of current LLMs are also constrained. Consequently, naive approaches that generate HDL code in a single conversation often lead to errors, as illustrated in Fig.~\ref{full} (a). One solution is to involve a human expert as a supervisor in the workflow, as seen in Chip-chat \cite{humanaid1}, where a human expert monitors the code generation process and provides instructions to correct the code, as illustrated in Fig.~\ref{full} (b). However, incorporating human experts increases the cost of the automatic design flow and may compromise the efficiency of the workflow.

ChipNeMo \cite{trained2} and other studies \cite{trained1,trained3,trained4,trained5} aim to enhance the quality of generated HDL code by fine-tuning LLMs with augmented datasets. This approach allows LLMs to learn more about HDL design, helping them to reduce the risk of generating incorrect HDL code. However, this method relies heavily on the quality of the augmented datasets and requires substantial computational resources for fine-tuning. As a result, fine-tuned models tend to be smaller than commercial models like GPT-4 \cite{gpt4}, which may limit their reasoning abilities.

Another method for improving the quality of automatically generated HDL code is to leverage the in-context learning capabilities of LLMs \cite{incontext}. This ability allows LLMs to learn how to generate new patterns based on examples provided in context. One application of this is retrieval-augmented generation (RAG) \cite{rag}, as illustrated in Fig.~\ref{full} (c). RAG retrieves examples from a database to guide code generation using LLMs. Previous work, such as GPT4AIGChip \cite{raghdl1, raghdl2, raghdl3}, uses RAG to either retrieve examples similar to specifications for generation assistance or to retrieve examples for error correction. However, these methods require a database with a large number of examples, which may necessitate human expertise to build. Additionally, if the database lacks relevant examples, retrieving unrelated ones may mislead the LLMs. AutoChip \cite{autochip} addresses this issue by using feedback from testbenches as a substitute for a database, thereby avoiding the need for database construction. However, in some scenarios, the final testbench only indicates the pass rate of test samples, which may not provide sufficient information for LLMs to correct errors in code generation.

To mitigate the hallucination problem of LLMs without relying on fine-tuning, human labor, databases, or feedback from testbenches, we introduce a human-expert-inspired method to enhance performance in HDL code generation. As illustrated in Fig.~\ref{full} (d), we first summarize design experiences into paradigm blocks, which contain prompts to guide the LLMs in breaking down problems according to human design methodology, along with scripts to call external tools. For each specification, LLMs classify the type of circuit to match it with a specified paradigm block. Depending on the circuit type, either the procedure for combinational logic (COMB) or the procedure for sequential logic (SEQU) is executed. In both procedures, LLMs first extract information from the specifications and generate an information list. This information list is then either transformed into a format suitable for processing by Electronic Design Automation (EDA) tools or utilized in a divide-and-conquer strategy. Finally, HDL code for the given specification is generated. Moreover, we implement a two-phase workflow for multi-round generation to effectively reuse intermediate results and achieve better outcomes within a limited number of generation and verification rounds. We summarize our contributions as follows:
\begin{itemize} 
    \item
    To mitigate the hallucination problem of LLMs, we design specialized paradigm blocks to break tasks into several sub-procedures that align with the design methodology of human experts. These sub-procedures are tailored to match the capabilities of the LLMs and the characteristics of the tasks.

    \item 
    We employ a two-phase workflow to enhance the quality of generated code within limited generation and verification rounds. In the first phase, specialized paradigm blocks are used to handle the specifications. In the second phase, a general paradigm block is employed to attempt HDL code generation, reusing intermediate results of higher quality.

    \item
    Experimental results demonstrate that our method significantly improves the functional correctness of the generated HDL code. The proposed method outperforms the baseline on the VerilogEval-human dataset \cite{veval}, achieving improvements of 4.7\%, 11.0\%, and 14.7\% in Pass@1, Pass@5, and Pass@10, respectively. In the VerilogEval-machine dataset, the proposed method also increases the pass rate by more than 5\% in both Pass@5 and Pass@10.
\end{itemize}

\section{Motivation}\label{sec:motivation}
To enhance the performance of large language models (LLMs) in hardware description language (HDL) code generation without the need for extensive training, it is crucial to leverage the in-context learning and reasoning abilities of these models. One approach involves introducing external information explicitly through retrieval-augmented generation (RAG) methods; however, this heavily relies on the quality of the databases, which can be costly to develop. Instead, our solution aims to feed external information implicitly by incorporating LLM-suited design procedures.

\begin{figure}[t!]
    \centering
    \includegraphics[width=0.48\textwidth]{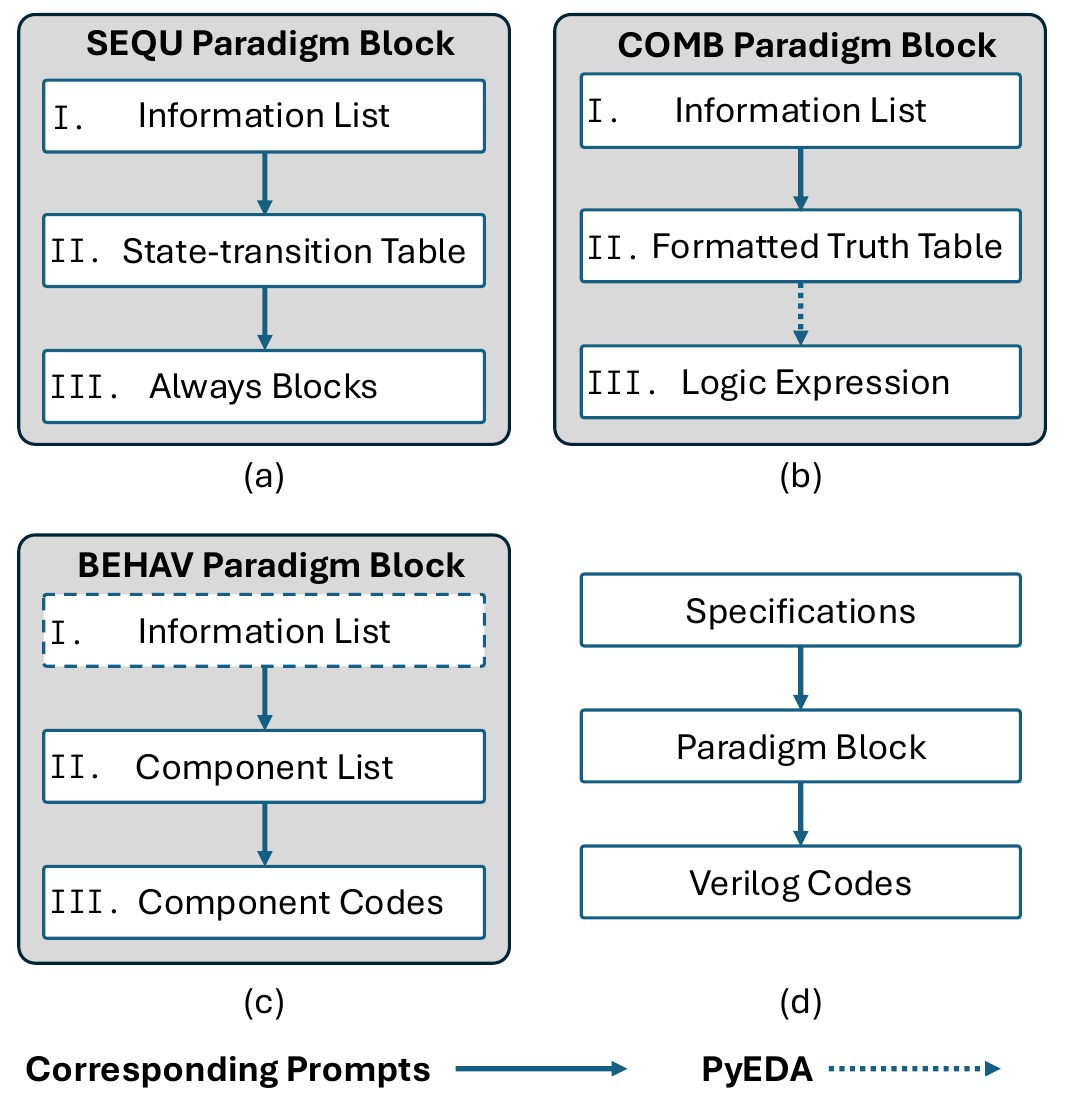}
    \caption{
    The proposed paradigm blocks SEQU, COMB, and BEHAV are illustrated in (a), (b), and (c), respectively. The overall procedure is outlined in (d). Solid arrows represent prompts to the LLMs, while the dashed arrow indicates the execution of PyEDA. 
    }
    \label{block}
\end{figure}

Unlike humans, a significant challenge for LLMs is their effectiveness in handling multi-step tasks, as they typically excel only at one-step tasks \cite{hop}. For instance, while LLMs can successfully retrieve a value B[A[x]] from arrays A[10] and B[10] in a single conversation, they often struggle with accessing C[B[A[x]]]. This limitation stems from the fact that state-of-the-art LLMs do not maintain sufficient internal states. To enhance their capability in managing multi-step tasks, it is crucial to store intermediate states within the prompts \cite{reasoning}.

To address this challenge, we simplify the HDL generation process by dividing it into several sub-tasks, with intermediate states between these sub-tasks transferred in an organized format. This approach aligns with the Chain-of-Thought (CoT) method \cite{cot}, which enhances LLM reasoning through self-planning within a single context. While CoT is effective for general-purpose question-answering, it is less suitable for HDL code generation due to the limited and detail-scarce training datasets available in the HDL design process. Furthermore, a single conversation is often insufficient for completing the design procedure. Drawing from the design practices of human experts, our strategy focuses on creating paradigms for different types of problems. These paradigms consist of multiple sub-tasks that fall within the reasoning capabilities of LLMs.

Another challenge is the inherent randomness in LLM generation. To avoid consistently arriving at sub-optimal solutions, tokens are randomly sampled based on the confidence of the LLMs. Consequently, there is always a probability of generating incorrect outputs \cite{randomness}. In scenarios where paradigms are formed by multiple sub-tasks, errors from earlier sub-tasks can propagate to later ones, increasing the overall error rate. To mitigate this impact, the paradigms should be compact, and it is necessary to utilize multi-round code generation while reusing the intermediate results.

\section{Proposed Approach}\label{sec:method}

\begin{figure}[t!]
{
\rule{\linewidth}{1.0pt}
% \vspace{0mm} % Add some space after the line
\textbf{\fontsize{10}{10pt}\selectfont Specification:}
}

{
    % \fontsize{10pt}{10pt}\selectfont
\textit{\fontsize{9pt}{9pt}\selectfont Build a decade counter that counts from 0 through 9,
inclusive, with a period of 10. The reset input is 
active high synchronous, and should reset the counter
to 0. We want to be able to pause the counter rather 
than always incrementing every clock cycle, so the 
"slowena" input if high indicates when the counter 
should increment.}\\
}
% \fontsize{10}{12pt}\selectfont
\textbf{\fontsize{10}{10pt}\selectfont Prompt: }\\
\textit{\fontsize{9pt}{9pt}\selectfont List the relationship between the inputs and outputs one by one with details as many as possible. Do not simplify! Keep everything as the same in the task. If it has reset and clk, list the timing requirements of reset and clk.\\}
\textbf{\fontsize{10}{10pt}\selectfont 
\rule{\linewidth}{0.4pt}
LLM Response:}\\\emph{\fontsize{9pt}{9pt}\selectfont
1. clk: This is the clock input to the module. The decade counter will increment based on this clock signal. The timing requirement for this signal would be that it must be periodic and consistent to ensure accurate counting.\\
...\\
4. q: This is the output of the module. It is a 4-bit signal that represents the current count of the decade counter. This signal will change at the rising edge of the clock if slowena is high and reset is low. If reset is high, q will be reset to 0 regardless of the state of slowena. \\
...\\
}
\rule{\linewidth}{1.0pt}
    \caption{Example of information list generated by LLMs for a sequential logic task. Some LLM responses are omitted due to space constraints.}

    \label{infolistsequapp}
\end{figure}

\begin{figure}[!h]
{
\rule{\linewidth}{1.0pt}
% \vspace{0mm} % Add some space after the line
\textbf{\fontsize{10}{10pt}\selectfont
Specification:}
\linespread{2.0} \selectfont
}

{

\textit{\fontsize{9pt}{9pt}\selectfont Implement the circuit described by the Karnaugh map below.}
\begin{verbatim}
//       ab
// cd   00 01 11 10
//  00 | 1 | 1 | 0 | 1 |
//  01 | 1 | 0 | 1 | 1 |
//  11 | 0 | 0 | 1 | 0 |
//  10 | 1 | 1 | 1 | 0 |
    \end{verbatim}
    }
\textbf{\fontsize{10}{10pt}\selectfont
Prompt:} \\
\textit{\fontsize{9pt}{9pt}\selectfont List the relationship between the inputs and outputs one by one with details as many as possible. Do not simplify! Keep everything as the same in the task. Do not write the codes! If it has reset and clk, list the timing requirements of reset and clk.\\}
\rule{\linewidth}{0.4pt}
\textbf{\fontsize{10}{10pt}\selectfont LLM:}\\
\emph{\fontsize{9pt}{9pt}\selectfont
Sure, here are the relationships between the inputs and output according to the Karnaugh map:\\
1. When a = 0, b = 0, c = 0 and d = 0, out = 1\\
...\\
16. When a = 1, b = 0, c = 1 and d = 0, out = 0\\
These are the complete relationships between the inputs and output as per the Karnaugh map, without any simplification. \\
}
\rule{\linewidth}{1pt}
    \caption{Example of information list generated by LLMs for a combinational logic task. Some LLM responses are omitted due to space constraints.}
    \label{infolistcombapp}
\end{figure}

In this section, we present a detailed overview of the proposed method. We begin by introducing the paradigm blocks, followed by an explanation of the overall two-phase workflow.

\subsection{Paradigm Blocks}
Human experts often rely on paradigms derived from their experiences to effectively tackle design problems. We simulate this process by creating paradigm blocks for large language models (LLMs). To maintain generality, we have designed three paradigm blocks for this work. Typical circuits include combinational and sequential logic; however, some specifications are difficult to classify into these two categories and are thus considered general. As illustrated in Fig.~\ref{block} (a) and (b), SEQU and COMB denote the paradigm blocks for sequential logic and combinational logic, respectively. BEHAV, as shown in Fig.~\ref{block} (c), serves as the general paradigm block to handle tasks that are challenging for SEQU or COMB. The information list in BEHAV is represented in a dashed box, indicating that it adopts information lists from the SEQU and COMB blocks. The overall procedure for generating Verilog code starts with input specifications, which are processed by a paradigm block to yield the generated Verilog code, as depicted in Fig.~\ref{block} (d).

As illustrated in Fig.~\ref{block} (a)-(c) I, the first step for all paradigm blocks is to generate the \textit{Information List}, referred to as information extraction. The purpose of this extraction is to transform the implicit information contained in the specifications into explicit information that directly outlines the relationships between module inputs and outputs.

The extraction process varies among the paradigm blocks. For combinational logic (COMB), the information list is structured as a truth table, while for sequential logic (SEQU), it is crucial to include timing descriptions alongside the relationships between inputs and outputs, such as whether input signals should be captured on the rising or falling edge. Examples of information lists are shown in \figname~\ref{infolistsequapp} and \figname~\ref{infolistcombapp}. The general paradigm block, BEHAV, is unique in that it reuses the information lists inherited from the COMB or SEQU blocks when it falls back to them.

After information extraction, the next step in the paradigm blocks is to convert the extracted information list into a standard format. For SEQU and COMB, as illustrated in Fig.~\ref{block} (a) II and (b) II, these formats are the state-transition table and the JSON-formatted truth table, respectively. Two examples of the generated state-transition table and the JSON-formatted truth table are shown in Fig.~\ref{sequproc} and Fig.~\ref{combproc}. The BEHAV block, as illustrated in Fig.~\ref{block} (c) II, being more general, focuses on decomposing tasks into various behavioral description components, such as representing a high bit-width adder with a \textit{for} loop. These components are organized into a component list with corresponding behavior descriptions. This structured approach naturally breaks down the input-output relationships described in the information lists into explicit and manageable elements, which reduces the reasoning load in subsequent steps. Additionally, the formatted data, such as the truth table, can be captured by scripts and processed by EDA tools. 

\begin{figure}[t!]
\rule{\linewidth}{1.0pt}
\textbf{\fontsize{10}{10pt}\selectfont
Information List:} ...\\
\textbf{\fontsize{10}{10pt}\selectfont Prompt:} \\
\textit{\fontsize{9}{9pt}\selectfont Give the state transition table with details as many as possible. Do not simplify! Keep everything as the same in the task. Do not give the verilog codes.}\\
\rule{\linewidth}{0.4pt}
\textbf{\fontsize{10}{10pt}\selectfont LLM:} \\
\textit{\fontsize{9}{9pt}\selectfont Sure, we can define a state machine with four states: IDLE, BYTE1, BYTE2, and BYTE3. Here's the state transition table:}\\
    \fontsize{8pt}{8pt}
    \begin{verbatim}
| Current State | in[3] | reset | Next State | done |
|---------------|-------|-------|------------|------|
| IDLE          | 0     | 0     | IDLE       | 0    |
| IDLE          | 0     | 1     | IDLE       | 0    |
| IDLE          | 1     | 0     | BYTE1      | 0    |
| IDLE          | 1     | 1     | IDLE       | 0    |
| BYTE1         | X     | 0     | BYTE2      | 0    |
| BYTE1         | X     | 1     | IDLE       | 0    |
| BYTE2         | X     | 0     | BYTE3      | 0    |
| BYTE2         | X     | 1     | IDLE       | 0    |
| BYTE3         | 0     | 0     | IDLE       | 1    |
| BYTE3         | 0     | 1     | IDLE       | 0    |
| BYTE3         | 1     | 0     | BYTE1      | 1    |
| BYTE3         | 1     | 1     | IDLE       | 0    |
    \end{verbatim}
\rule{\linewidth}{1.0pt}
    \caption{Example of the state-transition table generated by LLM for a sequential logic.}
    \label{sequproc}
\end{figure}

\begin{figure}[t!]
% \rule{\linewidth}{0.4pt}
\rule{\linewidth}{1.0pt}
\textbf{\fontsize{10}{10pt}\selectfont Information List: ...}\\
    \textbf{\fontsize{10}{10pt}\selectfont Prompt:} \\
    \textit{\fontsize{9}{9pt}\selectfont Give the truth table between the outputs and inputs in json. Do not simplify! Keep everything as the same in the task. Do not give the verilog codes. Use the json format like this example:}
    \fontsize{9pt}{9pt}
    \begin{verbatim}
{
    "table": [
    [0,0,0],
    [0,1,1],
    [1,0,1],
    [1,1,1]
    ],
    "inputs": ["a[1]","a[2]"],
    "outputs": ["x"]
    "header_inputs": ["a[2]","a[1]"],
    "header_outputs": ["x"]
}
    \end{verbatim}
\fontsize{10pt}{12pt}\selectfont
\rule{\linewidth}{0.4pt}
\textbf{\fontsize{10}{10pt}\selectfont LLM:} \\
\textit{\fontsize{9}{9pt}\selectfont Here is the truth table in json format}

\fontsize{9pt}{9pt}
    \begin{verbatim}
{
    "table": [
        [0,0,0,0,0], [0,1,0,0,1],
        [1,0,0,0,1], [1,1,0,0,1],
        [0,1,0,1,0], [0,0,0,1,0],
        [1,0,0,1,1], [1,1,0,1,1],
        [0,1,1,1,0], [0,0,1,1,1],
        [1,0,1,1,1], [1,1,1,1,1],
        [0,1,1,0,0], [0,0,1,0,1],
        [1,0,1,0,1], [1,1,1,0,1]
    ],
    "inputs": ["a","b","c","d"],
    "outputs": ["out"],
    "header_inputs": ["a","b","c","d"],
    "header_outputs": ["out"]
}
    \end{verbatim}
\rule{\linewidth}{1.0pt}
    \caption{Example of the truth table in JSON generated by LLM for a combinational logic, where an output example is given in prompt for LLM to generate the valid JSON results.}
    \label{combproc}
\end{figure}

\begin{figure}[t!]

\rule{\linewidth}{1pt}
\textit{\fontsize{9pt}{9pt}\selectfont
Output Logic: This block will determine the output Z based on the current state of the shift register and the inputs A, B, and C. The inputs A, B, and C will be used as select lines for the multiplexer. The output Z will be the selected bit from the shift register. This block will be combinational and will be sensitive to the current state of the shift register and the inputs A, B, and C.}

\fontsize{9pt}{9pt}\selectfont
\begin{verbatim}
always @(Q, A, B, C) begin
    case({A, B, C})
        3'b000: Z = Q[0];
        3'b001: Z = Q[1];
        3'b010: Z = Q[2];
        3'b011: Z = Q[3];
        3'b100: Z = Q[4];
        3'b101: Z = Q[5];
        3'b110: Z = Q[6];
        3'b111: Z = Q[7];
    endcase
end\end{verbatim}
\rule{\linewidth}{1pt}
\caption{Example of a \textit{always block} generated by LLM for task \textit{ece241\_2013\_q12}.}
\label{always}

\end{figure}

In the next step within the paradigm blocks, LLMs begin generating Verilog code. For SEQU, as illustrated in Fig.~\ref{block} (a) III, we employ the three-always-block method \cite{threephase}, instructing the LLMs to complete the \textit{always blocks} sequentially, and an example is shown in Fig. \ref{always}. For each \textit{always block}, we ask the LLMs to produce a description prior to generating the code, allowing for further reasoning. Subsequently, the LLMs merge the generated \textit{always blocks} into a complete module. For COMB, as illustrated in Fig.~\ref{block} (b) III, we leverage the external tool PyEDA to assist the LLMs. Performing calculations can be complex for LLMs, as they often provide answers that resemble the correct response rather than executing detailed calculations. Consequently, LLMs currently lack the ability to simplify truth tables accurately. Since the truth table is already formatted in JSON from the previous step, we utilize PyEDA to simplify it into a sum-of-products (SOP) expression. This method is faster and more accurate than relying solely on the LLMs. Following this, the LLMs generate the Verilog code based on the SOP expression. For BEHAV, as illustrated in Fig.~\ref{block} (c) III, the LLMs generate code for one component at a time according to the corresponding descriptions, and then integrate these component codes into a complete module.

\subsection{Two-phase Workflow}

\begin{figure}
    \centering
    \includegraphics[width=0.48\textwidth]{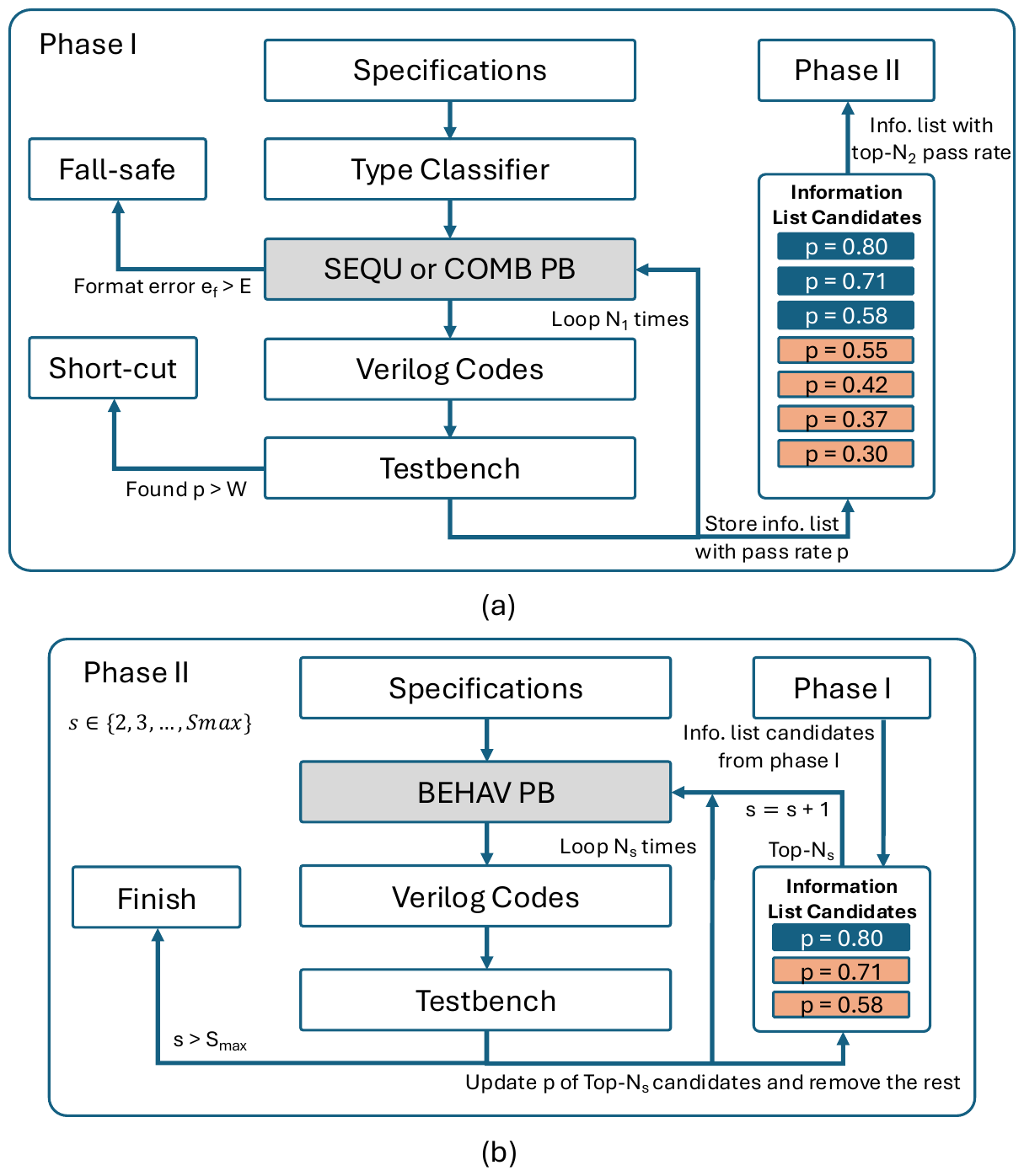}
    \caption{
    The proposed workflow begins with \textit{phase I}, as illustrated in (a), where specifications are classified and processed using either the SEQU or COMB paradigm block. Once \textit{phase I} is completed, \textit{phase II} commences, as shown in (b). In this phase, the information lists from \textit{phase I} are reused, and the BEHAV paradigm block is implemented.
    }
    \label{chain}
\end{figure}

Multi-round generation is an effective strategy for mitigating the inherent randomness in LLMs. However, executing the testbench an unlimited number of times is not feasible. The challenge, therefore, is to maximize the pass rate within a limited number of testbench executions, denoted as $N_{total}$. To address this challenge, we propose a two-phase workflow, as illustrated in \figname~\ref{chain}. The first phase focuses on generating code using the specialized SEQU or COMB paradigm blocks. If no generated sample passes the testbench, the second phase is initiated. In this phase, the general paradigm block, BEHAV, is employed while reusing the information list candidates generated in \textit{phase I}. The rationale for reusing the information lists is that their quality significantly influences subsequent steps, making it preferable to select high-quality information lists rather than to regenerate them.

At the start of \textit{phase I}, the first step involves classifying the specifications. As shown in \figname~\ref{chain} (a), the \textit{Type Classifier} is responsible for categorizing the specifications into SEQU or COMB. This classification is performed by LLMs. To mitigate errors caused by misleading semantics in specifications, for example, a design for combinational logic within a sequential logic context, we first instruct the LLMs to naïvely generate Verilog code for the target specification. The classification is then based on the generated code, which provides more context for the LLMs to accurately deduce the target circuit types, even if the resulting circuit may not be functionally correct at this stage.

After classification, the selected paradigm block enters a loop to generate $N_1$ code samples, where $N_1$ represents the allocated number of testbench execution times for \textit{phase I}. All code samples are verified using the testbench to obtain the pass rate $p = \frac{m}{n}$, where $n$ is the number of test cases in the testbench and $m$ is the number of cases that passed. In addition to indicating the quality of the generated code samples, the pass rate $p$ serves as a signature of the quality of the intermediate results. Consequently, the corresponding information list candidates are stored and ranked based on $p$. Upon completion of the loop, the top-$N_2$ information lists are forwarded to \textit{phase II}.

If \textit{phase I} concludes without yielding any passed code samples, \textit{phase II} commences. In \textit{phase II}, the BEHAV paradigm block is utilized for specifications that struggled with SEQU or COMB. Unlike the loop in \textit{phase I}, the loop in \textit{phase II} executes multiple times, denoted as $s \in \{2, 3,..., S_{max}\}$. For each loop, BEHAV and the testbench are executed $N_s$ times, using the top-$N_s$ of current information list candidates. Therefore, considering $N_1$ in \textit{phase I}, we have $N_{total} = \sum_{s=1}^{S_{max}} N_{s}$. After each loop, the new pass rates for the corresponding information lists are updated to the average of the last pass rate $p_{last}$ and the new pass rate $p_{new}$, denoted as $p = \frac{p_{last} + p_{new}}{2}$, and any unused information lists are discarded.

\begin{table*}[t!]
  \centering
    \caption{Comparison of Pass@$k$ for the Proposed Method and Baseline.}
  \begin{tabular}{c c c c c c c c c c c c c c c}
  \hline
  \toprule
  % \addlinespace[1pt]
  % \hline
  % \addlinespace[2pt]
   Pass@$k$ & & \multicolumn{6}{c}{VerilogEval-human} & & \multicolumn{6}{c}{VerilogEval-machine} \\
\cline {1-1} \cline {3-8} \cline {10-15}
\addlinespace[2pt]
    \% & & Baseline & FULL & COMB & SEQU & BEHAV & SYNTAX & & Baseline & FULL & COMB & SEQU & BEHAV & SYNTAX\\
    \hline
    \addlinespace[2pt]
    \multicolumn{15}{c}{\textbf{GPT-4}}\\
  \hline
  \addlinespace[2pt]
  Pass@1 & & 42.0 & \textbf{46.7} & 44.6 & 43.7 & 42.5 & 86.9 & & 63.8 & \textbf{65.2} & 63.8 & 65.1 & 63.9 & 92.9\\ 
  Pass@5 & & 58.6 & \textbf{69.6} & 62.2 & 63.9 & 60.7 & 98.8 & & 76.3 & \textbf{81.4} & 76.3 & 80.7 & 77.0 & 98.9\\ 
  Pass@10 & & 62.2 & \textbf{76.9} & 66.0 & 69.2 & 66.0 & 99.4 & & 78.3 & \textbf{85.3} & 78.3 & 83.9 & 79.7 & 99.3\\ 
  % \addlinespace[2pt]

  \hline
  \addlinespace[2pt]
  % \toprule\\
  \multicolumn{15}{c}{\textbf{GPT-4o-mini}}\\
  \hline
  % \toprule
  % \addlinespace[2pt]
  % \hline
  % \addlinespace[1pt]
  Pass@1 & & 55.1 & \textbf{68.2} & 63.0 & 66.0 & 61.4 & 90.9 & & 68.8 & \textbf{73.5} & 72.8 & 73.1 & 73.0 & 88.9\\ 
  Pass@5 & & 64.0 & \textbf{72.8} & 64.4 & 69.9 & 63.4 & 98.8 & & 74.7 & \textbf{77.8} & 75.3 & 76.5 & 75.9 & 95.3\\ 
  Pass@10 & & 66.7 & \textbf{74.4} & 64.7 & 71.2 & 64.1 & 100.0 & & 77.6 & \textbf{79.7} & 76.2 & 77.6 & 76.9 & 96.5\\ 
  % \addlinespace[2pt]

  \hline \toprule\\
  
  \end{tabular}

  \label{main}
\end{table*}

\subsection{\textit{Fail-safe} and \textit{Short-cut}}\label{sec:specialmode}
To address special situations in \textit{phase I}, we incorporate two strategies: \textit{Fail-safe} and \textit{Short-cut}, as depicted on the left side of Fig.~\ref{chain} (a). Due to the randomness of LLMs, \textit{format errors} may occur in the intermediate results of the paradigm blocks. In such cases, the scripts for the next step may fail to capture the necessary information. To handle this, each task is allocated $E_f$ retry chances for \textit{format errors}. During each retry, the current intermediate results are disregarded, and the previous step is rerun. If the number of format errors $e_f$ exceeds $E_f$, we conclude that the SEQU or COMB paradigm block is unsuitable for the specification. The process then enters \textit{Fail-safe} mode and switches to BEHAV, dedicating all remaining execution times $N_{left}$ to it. Unlike the standard \textit{phase II}, where information lists are reused, the information lists will be regenerated before executing BEHAV in this mode.

Conversely, if a generated code sample nearly passes the testbench, indicated by a pass rate $p$ exceeding a predefined threshold $W$, the process enters \textit{Short-cut} mode. In this case, we consider that the SEQU or COMB paradigm block is well-suited for the current specification and that a high-quality information list has been identified. Therefore, instead of progressing to \textit{phase II}, all remaining execution times $N_{left}$ are allocated to the current SEQU or COMB paradigm block, while the generation of new information lists is halted in favor of utilizing the promising one. 
\section{Experimental Results}\label{sec:results}
To demonstrate the performance of the proposed method, we conducted experiments on the VerilogEval \cite{veval} dataset, which contains 299 tasks, each with specifications, testbenches, and ground truth Verilog codes. This dataset is divided into two categories: VerilogEval-human and VerilogEval-machine. The specifications for the VerilogEval-human tasks are written by human experts, while the VerilogEval-machine tasks are generated by GPT-3.5-turbo \cite{gpt35} based on ground truth codes. For evaluation, we utilized GPT-4 \cite{gpt4} and GPT-4o-mini \cite{gpt4omini} within the proposed framework through OpenAI APIs. Despite the release of newer GPT-4 versions, we chose GPT-4-0613 as our primary model for analysis because it was released before the VerilogEval dataset was published, ensuring that its training dataset does not include this dataset. Additionally, we utilized the latest version of GPT-4o-mini (GPT-4o-mini-2024-07-18) to provide reference results from a smaller and newer model. The temperature of the LLMs was set to 0.5 for all experiments, and the maximum context length was configured to 4096. For simulation, we employed Icarus Verilog (iverilog) \cite{iverilog} to run the testbenches and the generated code.

We measured the functional correctness of the generated code using the Pass@$k$ \cite{passat} metric. In Pass@$k$, $k$ code samples are tested, and we evaluate the number of samples which pass the testbench. It is defined as:

\begin{equation}
    Pass@ k:=\underset{Problems }{\mathbb{E}}\left[1-\frac{\binom{n-c}{k}}{\binom{n}{k}}\right]
\end{equation}
where $n$ is the number of samples per task, and $c$ is the number of samples that passed the testbench. In our functional correctness comparison, the baseline method involves LLMs directly generating code from specifications in a single conversation. Here, we consider each execution of the testbench as a code sample tested in Pass@$k$. We set $n$ to 10 and $k$ to 1, 5, and 10 for the experiments, respectively. The maximum format retry $E_f$ was set to 10 for both the proposed method and the baseline. For the search parameters, we set the maximum iteration $S_{max}$ to 3. Thus, the configuration for the number of generated code samples can be expressed as $(N_1, N_2, N_3)$, which is set to $(7, 2, 1)$. The \textit{Short-cut} threshold \textit{W} is set to 0.95. To reduce token costs, we first conducted the baseline experiments and then selected the hard tasks, for which the baseline method could not generate valid code to pass the testbench after 10 retries. The hard tasks are then processed by the proposed framework.

\begin{figure}[t!]
  \centering
  \includegraphics[width=0.48\textwidth]{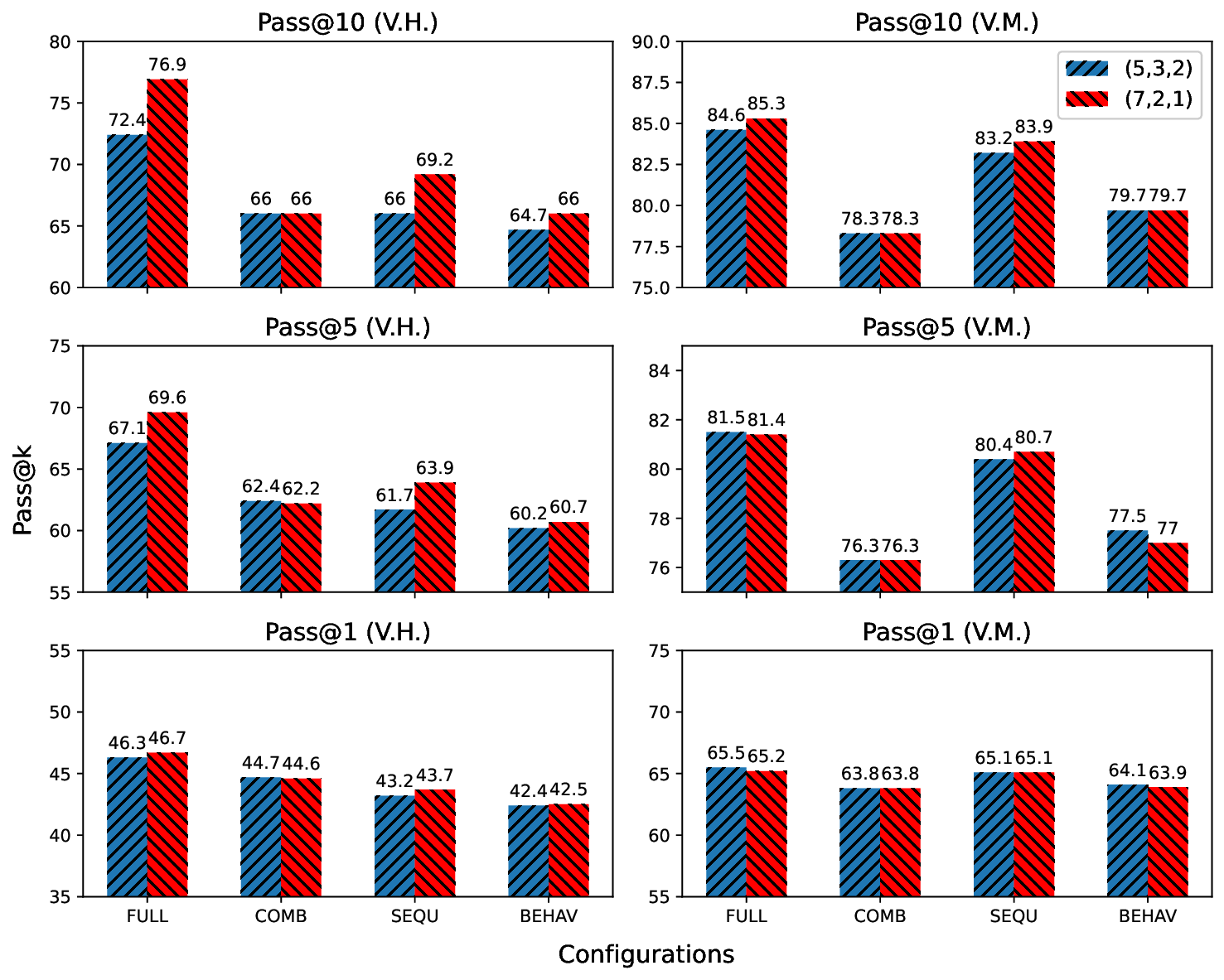}
  \caption{Comparison of the Pass@k under GPT-4 between configurations, (5,3,2) and (7,2,1), where V.H. and V.M. represent VerilogEval-human and VerilogEval-machine. The total results of all the procedures are shown as FULL, while the individual results of the COMB, SEQU and BEHAV procedures are shown as COMB, SEQU, and BEHAV, respectively.}
  \label{configcompare}
\end{figure}

As shown in Table \ref{main}, for GPT-4, the proposed method outperforms the baseline in VerilogEval-human with improvements of 4.7\%, 11.0\%, and 14.7\% in Pass@1, Pass@5, and Pass@10, respectively, and shows over 5\% improvement in Pass@5 and Pass@10 for VerilogEval-machine. Notably, the COMB procedure did not show any improvement in the VerilogEval-machine dataset. This is because the specifications in VerilogEval-machine are generated by summarizing the ground truth codes with LLMs, which already exhibit the logical relationships necessary for solving combinational logic tasks, allowing LLMs to bypass calculations and reasoning. For GPT-4o-mini, the proposed method outperforms the baseline in VerilogEval-human with improvements of 13.1\%, 8.8\%, and 7.7\% in Pass@1, Pass@5, and Pass@10, respectively, and shows over 3\% improvement in Pass@1 and Pass@5 for VerilogEval-machine. Although GPT-4o-mini is significantly smaller than GPT-4, its baseline outperforms GPT-4 in all Pass@$k$ metrics for VerilogEval-human and in Pass@1 for VerilogEval-machine. In proposed method, GPT-4o-mini also surpasses GPT-4 in Pass@1 and Pass@5 for VerilogEval-human and in Pass@1 for VerilogEval-machine, suggesting that VerilogEval may be included in the training datasets of GPT-4o-mini.

\begin{figure}[t!]
  \centering
  \includegraphics[width=0.48\textwidth]{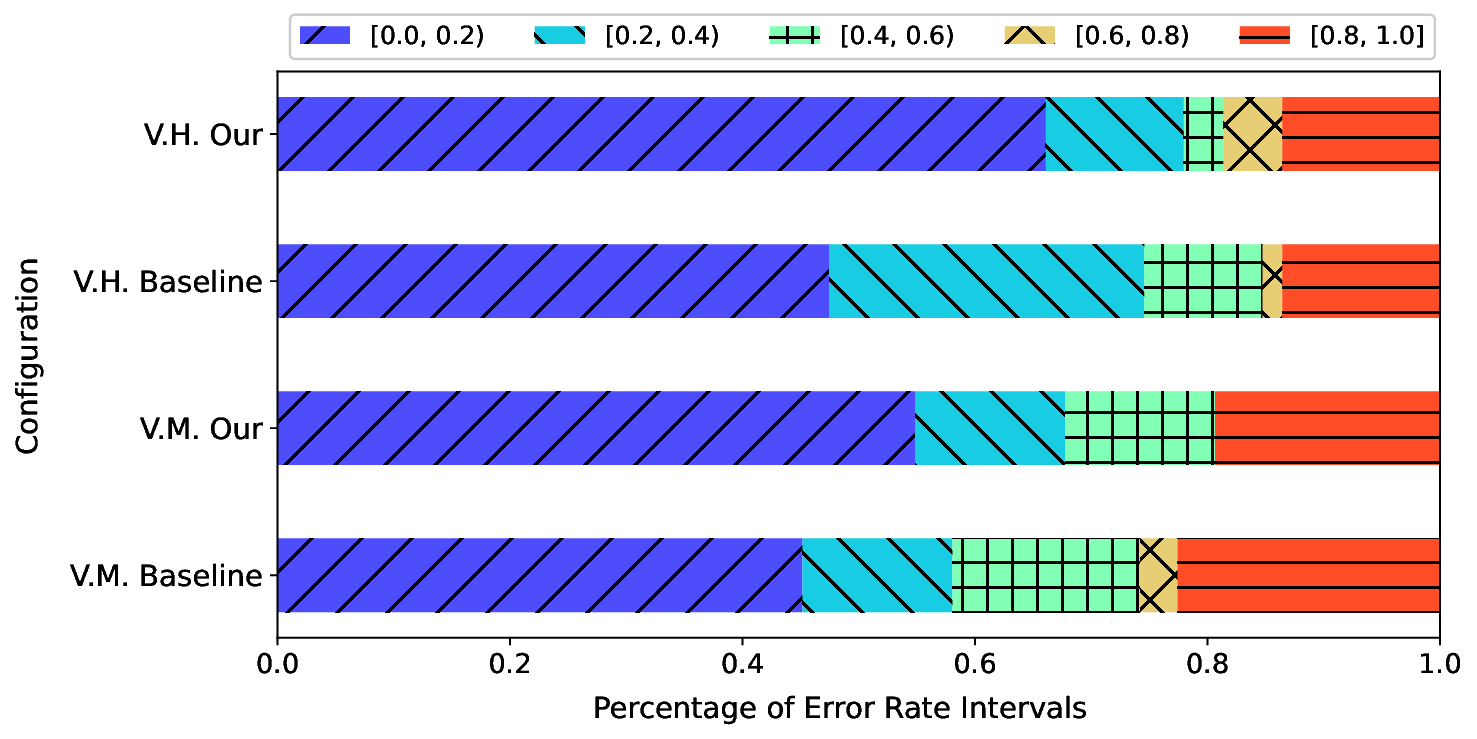}
  \caption{Comparison of testbench error rate distributions of GPT-4 for hard tasks. The x-axis represents the percentage of error rate intervals, and the y-axis represents the configurations. V.H. and V.M. represent VerilogEval-human and VerilogEval-machine.}
  \label{distribution}
\end{figure}

To demonstrate the influence of the search parameters, we compared the results of the search configurations $(5, 3, 2)$ and $(7, 2, 1)$ for GPT-4 in Fig.~\ref{configcompare}. We observe that the configuration $(7, 2, 1)$ outperforms $(5, 3, 2)$ in most cases, except for Pass@1 and Pass@5 in VerilogEval-machine. Overall, the configuration $(7, 2, 1)$ is superior to $(5, 3, 2)$ in the VerilogEval dataset. A potential reason for this is that configuration $(7, 2, 1)$ provides more information list candidates, increasing the likelihood of identifying a high-quality information list.

The testbench error rate distributions of GPT-4 for the baseline and the proposed method in the hard tasks of VerilogEval are illustrated in Fig.~\ref{distribution}, where the testbench error rate for each task is based on the best code sample in Pass@10. We divided the error rate distributions into five intervals with a step of 0.2. The proportions of the intervals are shown on the x-axis. It is evident that the proposed method not only improves Pass@$k$ but also reduces the error rates of code samples. However, the number of the hardest tasks, those with error rates exceeding 0.8, as indicated by the rightmost bars, did not decrease in VerilogEval-human and only slightly decreased in VerilogEval-machine. This suggests that there remains room for improvement in the proposed method.

\begin{table}[t!]
  \centering
    \caption{Comparison of AutoChip and the Proposed Method}
  \begin{tabular}{c c c}
  \hline
  \toprule
   \% & AutoChip & Our method \\
  \toprule
  \addlinespace[2pt]
  Sequential logic & 40.0 & 50.0 \\
  Combinational logic & 0.0 & 70.0 \\
  \hline \toprule\\
  
  \end{tabular}

  \label{autochip}
\end{table}

\begin{table}[!t]
  \centering
    \caption{Comparison of Type Classification Accuracy}
  \begin{tabular}{c c c}
  \hline
  \toprule
   \% & w/ code & w/o code \\
  \toprule
  \addlinespace[2pt]
  VerilogEval-human & 98.3 & 94.9 \\
  VerilogEval-machine & 100.0 & 100.0 \\
  \hline \toprule\\
  
  \end{tabular}

  \label{type}
\end{table}

As fine-tuning methods, such as those in \cite{trained1}, primarily focus on GPT-3.5, and methods using RAG depend on additional database, here, we compare the performance of AutoChip \cite{autochip}. In Table \ref{autochip}, we present a comparison of the performance of AutoChip and the proposed method under GPT-4. For both sequential and combinational logic, we randomly selected ten tasks from the hard tasks in the VerilogEval-human dataset. To evaluate the performance of the methods under the same testing standard, we adhered to the Pass@$10$ standard, where each execution of the testbench is counted as a test, regardless of whether it occurred within an iteration or after a complete attempt of the method. For AutoChip, the feedback iteration $n$ was set to 10. The results indicate that the proposed method outperforms AutoChip, particularly in combinational logic. This is because AutoChip heavily relies on testbench feedback, which is limited in the VerilogEval dataset. In contrast, the proposed method operates independently of testbench feedback, effectively addressing the weaknesses of LLMs in combinational logic.

\begin{figure}[t!]
\rule{\linewidth}{1.0pt}
% \vspace{0mm} % Add some space after the line
{\fontsize{9}{9pt}\selectfont\textbf{Task (a) Passed Case:}}\\
\emph{\fontsize{9}{9pt}\selectfont
q: This is a 64-bit register which holds the contents of the shift register. It gets updated with the new value either when the shift operation takes place (if ena is high) or when the new data is loaded into the shift register (if load is high). The updating of this register takes place on the positive edge of the clk signal. } 
\rule{\linewidth}{0.4pt}
{\fontsize{9}{9pt}\selectfont\textbf{Task (a) Failed Case:}}\\
\emph{\fontsize{9}{9pt}\selectfont
q: This is the 64-bit output of the shift register. It represents the current contents of the shift register. }

% \small
\rule{\linewidth}{1.0pt}
% \vspace{0mm} % Add some space after the line
{\fontsize{9}{9pt}\selectfont\textbf{Task (b) Passed Case}}\\
\emph{\fontsize{9}{9pt}\selectfont
When a=0, b=0, c=1, d=0, the output is 1.\\
When a=1, b=0, c=1, d=0, the output is 1. \\}
\rule{\linewidth}{0.4pt}
{\fontsize{9}{9pt}\selectfont\textbf{Task (b) Failed Case}}\\
\emph{\fontsize{9}{9pt}\selectfont
When a=0, b=0, c=1, d=0, the output is 0.\\
When a=1, b=0, c=1, d=0, the output can be 0 or 1 (don't care). }
\rule{\linewidth}{1.0pt}

  \caption{Comparison of snippets of information lists between two attempts of generation. For \textit{Task (a)} and \textit{Task (b)}, above is from the case passed the testbench, and below is from the case failed in the testbench.}
  \label{infolistres}
\end{figure}

Table \ref{type} presents a comparison of type classification accuracy in GPT-4 with and without the use of the generated codes. In the VerilogEval-machine dataset, both methods correctly classify the type of the circuit. However, in the VerilogEval-human dataset, the method using codes generated by LLMs has a 1.7\% error rate, while the naive method has a 5.1\% error rate. An analysis of the error cases reveals that these errors are due to misleading specifications that require combinational logic components from a sequential logic circuit. The naive method focuses more on the semantics of the specifications, while the method using codes generated by LLMs can leverage internal knowledge of Verilog design to avoid some errors.

In Fig.~\ref{infolistres}, the information lists from two tasks are compared. For each task, one information list is from the code that passed the simulation and the other is from the code that failed. In the case of the sequential logic task, as shown in Fig.~\ref{infolistres} \textit{Task (a)}, the information list of the failed code contains fewer descriptions compared to the list of the code that passed the simulation. Due to the randomness in LLMs, some information lists may be generated with information loss, leading to the generation of failed code. While detailed descriptions might be considered redundant by human experts, they help the LLMs focus on key information. In the examples of the combinational logic task, as shown in Fig.~\ref{infolistres} \textit{Task (b)}, the information list that failed in the simulation contains an incorrect output description. Both outputs, (a=0, b=0, c=1, d=0) and (a=1, b=0, c=1, d=0), are incorrect, which is caused by the hallucination of the LLMs.
\section{Conclusion}
\label{sec:conclusion}

In this work, we proposed a human-expert-inspired method to enhance the performance of HDL code generation using LLMs. First, we constructed specialized paradigm blocks that consist of several steps designed to divide and conquer generation tasks. Specifications are matched with the appropriate paradigm block and execute the block to generate the HDL code. Additionally, we introduced a two-phase workflow for multi-round generation, aimed at effectively improving the testbench pass rate of the generated HDL code within a limited number of generation and verification rounds. The experimental results demonstrate that our method significantly improves the functional correctness of the generated HDL.

\let\oldbibliography\thebibliography
\renewcommand{\thebibliography}[1]{%
\oldbibliography{#1}%
\fontsize{7.8pt}{7.8}\selectfont
\setlength{\itemsep}{0.5pt}%
}

% \clearpage
\bibliographystyle{IEEEtran}
\bibliography{bibfile}

\end{document}